\documentclass[12pt,preprint]{aastex}
\usepackage[dvipdfm]{hyperref}

\newcommand {\ea} {{\it et~al.}}
\newcommand {\be} {\begin{equation}}
\newcommand {\ee} {\end{equation}}
\newcommand {\gi} {\bar\gamma_{\rm inj}}

\shorttitle{The origin of X-ray spectra in blazars}

\shortauthors{Sikora \ea}

\begin{document} 

\title{On the origin of X-ray spectra in luminous blazars}

\author{Marek~Sikora\altaffilmark{1,4}, 
Mateusz~Janiak\altaffilmark{1}, 
Krzysztof~Nalewajko\altaffilmark{2,3}, 
Greg~M.~Madejski\altaffilmark{4}, 
and Rafa{\l}~Moderski\altaffilmark{1}} 

\altaffiltext{1}{Nicolaus Copernicus Astronomical Center, Bartycka 18, 
  00-716 Warsaw, Poland; {\tt sikora@camk.edu.pl}, {\tt mjaniak@camk.edu.pl}} 
\altaffiltext{2}{JILA, University of Colorado and National Institute of 
  Standards and Technology, UCB 440, Boulder, CO 80309, USA} 
\altaffiltext{3}{NASA Einstein Postdoctoral Fellow} 
\altaffiltext{4}{Kavli Institute for Particle Astrophysics and 
  Cosmology, SLAC National Accelerator Laboratory, Stanford 
  University, 2575 Sand Hill Road M/S 29, Menlo Park, CA 94025, USA}

\begin{abstract}
Gamma-ray luminosities of some quasar-associated blazars imply  jet powers 
reaching values comparable to the accretion power even if assuming very 
strong Doppler boosting and very high efficiency of gamma-ray production.
With much lower radiative efficiencies of protons than of electrons,
and the recent reports of very strong coupling of electrons with shock-heated 
protons indicated by Particle-in-Cell (PIC) simulations, the leptonic models 
seem to be strongly favored over the hadronic ones.  However, the electron-proton 
coupling combined with the  ERC (External-Radiation-Compton) models of gamma-ray 
production in leptonic models predict extremely hard X-ray spectra, with energy 
indices $\alpha_x \sim 0$.This is inconsistent with the observed 2-10 keV slopes 
of blazars, which cluster around $\alpha_x \sim 0.6$.  This problem can be 
resolved by assuming that electrons can be efficiently cooled down radiatively to 
non-relativistic energies, or that blazar spectra are entirely dominated by the SSC 
(Synchrotron-Self Compton) component up to at least 10 keV.  
Here, we show that the required cooling can be sufficiently efficient only at 
distances $r < 0.03$ pc.  SSC spectra, on the other hand, can be 
produced roughly co-spatially with the observed synchrotron and ERC components, which 
are most likely located roughly at a parsec scale.  We show that the dominant SSC component 
can also be produced much further than the dominant synchrotron and ERC components, at 
distances of $\gtrsim 10$ parsecs.  Hence, depending on the spatial distribution of the 
energy dissipation along the jet, one may expect to see $\gamma$-ray/optical events 
with either correlated or uncorrelated X-rays.  In all cases the number of e$^+$e$^-$ 
pairs per proton is predicted to be very low.  The direct verification of the proposed 
SSC scenario, and particularly the question of the co-spatiality of the SSC component 
with other spectral components, requires sensitive observations in the hard X-ray band.  
This is now possible with the deployment of the NuSTAR satellite, providing the required 
sentitivity to monitor the details of the hard X-ray spectra of blazars in the range 
where the ERC component is predicted to start dominating over the SSC component.  
\end{abstract}


\section{Introduction}

Images of extended  jets in radio-galaxies and quasars 
show that jet energy is dissipated more or less smoothly 
over all spatial scales. But in powerful, FR~II type radio sources, 
a large fraction of energy is very efficiently transmitted up to hundreds of 
kiloparsecs and dissipated there in terminal shocks. 
Studies of energy content of radio lobes indicate that they are powered by jets 
at rates sometimes comparable or even exceeding the accretion power 
(Rawlings \& Saunders 1991; Punsly 2007; Fernandes et al. 2011). 
Such extreme energetics is independently confirmed by studies of luminous  
blazars (Ghisellini et al. 2010; Ghisellini et al. 2011). These objects, with 
relativistically boosted jets pointing almost exactly at us, allow tracing 
the jet structure at parsec/subparsec distances from the black 
hole. Their structure on such scales is explored by multiwavelength 
studies of variability. However, the multi-band time series of blazars 
are complex, precluding a consensus regarding the physics of AGN jets --- 
their power, matter content, magnetization, cross-sectional structure, etc.  
This is not surprising, given that on such scales a variety of processes 
may contribute to the jet evolution and its nonthermal activity.  
Presumably, a conversion from the magnetic to the matter energy flux 
dominated flow takes place already at subparsec scales (Sikora et al. 2005).  
This conversion could be triggered by MHD instabilities, and governed 
by efficiency of the magnetic reconnection (Begelman 1998; Giannios \& 
Spruit 2006; Lyubarsky 2010).  Non-steady and non-axisymmetric 
jet launching, as is predicted by  the scenario  which involves MCAF 
(Magnetically-Choked Accretion Flows; McKinney et al. 2012), may strongly amplify 
these processes and generate strong internal shocks (Spada et al. 2001).   
Finally, due to interaction of the flow with external medium, 
oblique/reconfinement shocks are expected to be formed (Daly \& Marscher 1988; 
Komissarov \& Falle 1997; Nalewajko \& Sikora 2009). 

Given the complexity of the jet structure, one might expect that dissipation 
processes in blazars are not limited to a single zone, but rather they 
operate independently over two or more sites at once, and with different 
and possibly variable efficiencies.  Hence, since different radiation 
spectra are produced at different sites, one might expect a broad range 
of correlations and time lags between different spectral bands (see, e.g., 
Janiak et al. 2012).  However, attempts to use multiwavelength 
observations to associate the specific spectral portions with a given dissipation 
site are still hampered by insufficient models of particle acceleration, 
particularly regarding the behavior electrons in the presence of ions.  
Electrons need to tap a significant fraction of dissipated energy in order 
to explain large luminosities of blazars, otherwise this energy would go to 
protons which only under very specific conditions can radiate 
efficiently, or can efficiently trigger processes leading to the production 
of secondary electrons/positrons (Sikora 2011).  A variety of mechanisms were 
suggested to preheat electrons up to the thermal level of the shocked 
protons, and allow them to participate in the diffusive shock acceleration 
process (e.g. Hoshino et al. 1992; Hoshino \& Shimada 2002).  Recent Particle-In-Cell 
(PIC) simulations demonstrated strong electron-proton coupling in shocks 
(Sironi \& Spitkovsky 2011), and thus confirmed the expected potential of 
the leptonic models to generate very luminous events in blazars.  

In order to map the structure of nuclear jets in quasars, it is also necessary 
to know the geometry of external radiation sources, which provide seed 
photons for the ERC production of $\gamma$-rays.  At least the structures 
responsible for broad emission lines are expected to be stratified and 
flattened (see Wills \& Brown 1986; Krolik et al. 1991; Horne et al. 1991; 
Arav et al. 1998; Gaskell et al. 2007; Czerny \& Hryniewicz 2011).  
Both strong proton-electron coupling and such geometries are critical 
ingredients in our approach to establish the sites of the observed radiation 
spectra in luminous blazars.  

This paper is organized as follows:  
in Section \ref{sec_jetpower}, we discuss the implications of very large 
$\gamma$-ray luminosities and of strong electron-proton coupling for 
radiative scenarios;  Section \ref{sec_efficiency} formulates the 
connection of the energy dissipation efficiency with the average electron 
injection energy and e$^+$e$^-$-pair content;  in Section 
\ref{sec_consequencies}, we investigate possible mechanismsof X-ray
production in luminous blazars in light of the strong electron-proton coupling 
and the large electron injection energy.
Our main results are discussed in Section \ref{sec_disc} and summarized in 
Section \ref{sec_concl}.  

\section{Jet powers in luminous $\gamma$-ray blazars} 
\label{sec_jetpower} 

The radiative output of luminous blazars  associated with  
FSRQs (Flat-Spectrum Radio Quasars) is often strongly dominated 
by $\gamma$-rays. For observers located at an angle $1/\Gamma$ 
to the jet axis, the apparent $\gamma$-ray luminosity of such objects is 
\be
L_{\gamma} \simeq \eta_{\gamma} \eta_{p/e} \eta_{\rm diss} \, \Gamma^2 L_{j,0}\,,
\ee
where $\eta_{\rm diss}$ is the overall dissipation efficiency, 
$\eta_{p/e}$ is the fraction of dissipated energy channeled to accelerated 
protons or electrons, $\eta_{\gamma}$ is the fraction of energy of 
the accelerated 
particles emitted in the $\gamma$-ray band and $L_{j,0}$ is the jet power 
before dissipation region. Depending on whether $\gamma$-rays 
are produced by directly accelerated/heated electrons (leptonic models) 
or by protons 
and products of their interactions with photons and/or matter (hadronic 
models), $\eta_{p/e} = \eta_e$ or $\eta_p$, respectively.  

Noting that the maximal jet power is limited by the accretion power 
(McKinney et al. 2012 and refs. therein):
\be
\frac {L_{j,0}}{\dot M_d c^2} = 
\frac {L_{\gamma} \epsilon_d}{\eta_{\gamma} \eta_{p/e} \eta_{\rm diss} \Gamma^2 L_d}  
\, \lesssim 1 \,,
\ee
which gives
\be \eta_{\gamma}\eta_{p/e}\eta_{\rm diss} \gtrsim 
\frac {1}{4} \frac {L_{\gamma,49} (\epsilon_d/0.3)}
{(\Gamma/20)^2 (L_d/0.3 L_{\rm Edd}) M_{\rm BH,9}} 
\, , \ee
where $\dot M_d$ is the disk accretion rate,
$L_d$ is the accretion disk luminosity,
$\epsilon_d$ is the disk radiative efficiency,
and $M_{\rm BH,9} = M_{\rm BH}/10^9 M_{\odot}$. 
This means that \emph{all efficiencies must be high}. However there are certain 
constraints on some of them. Very demanding energetics of extended, FR~II 
radio sources (Rawlings \& Saunders 1991; Punsly 2007; Fernandes et al. 2011) 
indicates that the jet cannot lose most of its energy before reaching the 
terminal shocks in hot spots, therefore, $\eta_{\rm diss}$ 
is expected to be less than $\sim 0.5$.  
Even stronger constraints on $\eta_{\rm diss}$ are provided by 
models of reconfinement/oblique shocks (Nalewajko 2012) as well as internal, 
relativistically propagating shocks (Spada et al. 2001).  
Regarding $\eta_p$ no severe constraints exist, 
at least in the shock models. Likewise, given a very strong coupling 
between protons and electrons in shocked plasmas, 
as indicated by PIC simulations (Sironi \& Spitkovsky 2011), 
no severe constraints are imposed on $\eta_e$.  
For $n_e = n_p$, they can share their total energy equally, 
e.g. $\eta_e \simeq \eta_p \simeq 0.5$.  
Efficiency of the gamma-ray production, 
$\eta_{\gamma}$, may have a very broad range depending on a distance from 
the black hole and on the particle injection spectrum.  
Efficient cooling of protons is possible only if they are injected very 
close to the black hole, at $r < 100 R_g$, and if most of them are injected 
with ultrarelativistic energies (Sikora 2011). In the case of the power-law 
injection the latter condition implies the injection spectral index $p<1$.  

In leptonic models, $\gamma$-rays can be efficiently produced by relativistic 
electrons up to several parsecs.  
Because of the strong electron-proton coupling, electrons are 
preheated up to relativistic energies with the quasi-Maxwellian distribution 
with similar average energy as protons.  The Maxwellian distribution 
naturally explains the formation of very hard low-energy tail of injected electrons.  
All the above favors the leptonic radiative models, and only such will be investigated below. 

\section{Energy dissipation efficiency and average electron injection energy} 
\label{sec_efficiency} 

Blazars can be powered by kinetic energy of cold protons, as well as by various forms of 
internal energy -- magnetic or macro-turbulent. The kinetic one can be 
dissipated via the 
reconfinement/oblique shocks (Daly \& Marscher 1988; Komissarov \& Falle 1997; 
Nalewajko \& Sikora 2009), the macro-turbulent one --  via the 
internal shocks (Spada et al. 2001), and the magnetic one --  via 
the reconnection (Lovelace, Newman \& Romanova 1997; Lyubarsky 2010;  
Nalewajko et al. 2011). While 
the dissipation of jet energy in a reconfinement shock does not involve 
motion of the dissipation sites, internal shocks and reconnection layers 
form a sequence of moving sites. However, noting that the blazar 
high states, albeit very variable, last usually much longer than 
the time scale of passing of the flow through the distance range where most 
of the blazar radiation is produced, one may approximate the dissipation 
zone as steady-state in all cases.  

Assuming $n_e/n_p \ll m_p/m_e$ one can write the jet power in the form  
$L_j = L_p + L_{\rm int}$,
where $L_p$ is the flux of kinetic energies of cold protons, 
\be L_p = (dN_p/dt)  m_p c^2 (\Gamma-1)  \, , \label{Lp} \ee
and $L_{\rm int}$ is the  flux of internal energies. For the conserved 
proton number flux, $dN_p/dt = {\rm const}$, that provides formula 
for an  efficiency of the jet energy dissipation within a given region 
\be
\eta_{\rm diss} = \frac {(L_{j,0} - L_j)}{L_{j,0}} = 
1 - \frac {(\Gamma-1)(1+\sigma)}{(\Gamma_0-1)(1+\sigma_0)}\,,
\label{diss} 
\ee
where $\sigma=L_{\rm int}/L_p$, $\Gamma$ is the jet Lorentz factor, and quantities 
with the subscript '0' are the initial values of the variable / parameter.

\subsection{Average energy of injected electrons}
\label{sec_injenergy}

Electrons tap energy at a rate
\be
\frac{dN_{e,\rm inj}}{dt} \gi\,m_ec^2 \Gamma =
\eta_e \eta_{\rm diss} L_{j,0} \, ,
\label{inj1}
\ee
where $\gi m_ec^2 \Gamma$ is the average energy gained by 
an electron and 
$dN_{e,\rm inj}/dt = \int_{\gamma_{\rm min}}^{\gamma_{\rm max}} {Q_{\gamma}\,{\rm d}\gamma}$ is 
the electron injection rate. Assuming that each electron is accelerated 
once gives
\be \frac{dN_{e,inj}}{dt} = \frac{dN_e}{dt} = \frac{n_e}{n_p}\,  \frac{dN_p}{dt} 
\, , \label{inj2} \ee
and combining Eqs.(\ref{inj1}), (\ref{inj2}) and (\ref{Lp}) we obtain that for
$\Gamma \gg 1$:
\be
\gi = \frac{n_p m_p}{n_e m_e} \, \frac{\eta_e\eta_{\rm diss}}
{(1-\eta_{\rm diss})} \, (1+\sigma)\,.
\label{inj}
\ee
%

\subsection{Pair content}
\label{sec_pairs}

By modeling the blazar spectra, one can estimate the value of $\gi$,
and using Eq. (\ref{inj}) one may estimate the e$^+$e$^-$-pair content:
\be
\frac{n_e}{n_p} = \left(\frac{m_p}{m_e}\right)
\left(\frac{\eta_{\rm diss}}{1-\eta_{\rm diss}}\right)
\,\frac{(\eta_e/0.5)(1+\sigma)}{\gi}\,,
\label{pair}
\ee 
For $\sigma <1$, $\eta_{\rm diss} < 0.5$, and $\eta_e=0.5$ (implied by the 
strong electron-proton coupling), Eq.~(\ref{pair}) 
\\
gives
$n_e/n_p < 10$ for $\gi = 100$, and $n_e \sim n_p$ for $\gi =500$.  
Similar constraint on a pair content has been derived, but using different
arguments, by Ghisellini \& Tavecchio (2012).~\footnote{Note that much larger pair
content, predicted by Sikora \& Madejski (2000), 
was obtained assuming that X-rays are contributed by the low energy 
tail of the ERC spectral component produced by electrons injected with much 
lower average energy than resulting from the proton-electron coupling.}

\section{Spectral consequences of the strong proton-electron coupling}
\label{sec_consequencies}

Strong coupling between protons and electrons implies 
the break in the electron injection spectrum just below $\gi$ and formation
of an extremely hard low-energy tail, with the index
$p<1$ ($Q_{\gamma} \propto \gamma^{-p}$). In the slow-cooling
regime, i.e. where the electron energy losses are dominated by adiabatic losses,
such an injection function leads to the low portion of the
electron-energy-distribution
with the slope $s=1$ ($N_{\gamma} \propto \gamma^{-s}$), and radiation flux
index $\alpha = 0$ ($F_{\nu} \propto \nu^{-\alpha}$). 
Since typically the  soft/mid X-ray spectra in luminous blazars have 
much softer slopes ($\alpha_x \sim 0.6$: Abdo et al. 2010a; Ghisellini et al.
2011; Giommi et al. 2012),
they cannot be low energy tails of the ERC spectral component if produced in 
the slow cooling regime. 
One can exclude also production of such spectra by superposition 
of the SSC component with the very hard low-energy ERC component, because that 
would require fine tuning of model parameters, particularly 
if the SSC spectrum in the X-ray band has a slope $\alpha >1$.  
But there are still two other options: (i) production of 
soft/mid X-rays by the ERC process in the fast cooling regime; (ii) the SSC 
process with the luminosity peak at $h\nu_x \gtrsim 30$ keV.  
Both are examined below.

\subsection{X-ray spectra as the low energy tail of the ERC component 
in the fast cooling regime?}
\label{sec_tail}

Due to efficient radiative losses, the low-energy portions of the ERC spectra 
extend down to $h\nu_c \simeq \gamma_c^2\Gamma^2 \nu_{\rm ext}$ with the slope 
$\alpha \simeq 0.5$, where
\be
\gamma_c \simeq \frac {m_ec^2} {\sigma_T}\,  
\frac {\Gamma}{r u_{\rm ext}'}
\label{gammac}
\ee
is the cooling break, and
\be
u_{\rm ext}' = {g_u \xi L_d \Gamma^2 \over 4 \pi r^2 c}\,,
\label{uext}
\ee
is the energy density of the external radiation in the jet co-moving frame,
$\xi$ is the fraction of the accretion disc luminosity, $L_d$,
reprocessed in the BLR and dusty torus, and $g_u$ is the numerical factor 
which depends on the geometry of the external radiation sources. 
For stratified and flattened source geometries the value of $g_u$ is expected 
to be of the order of $0.1$  (see Appendix A and Fig. A1). 
Since for $\gamma < \gamma_c$ the adiabatic 
losses start to dominate and the spectrum breaks down to $\alpha \sim 0$, 
in order to explain the much softer observed X-ray spectra, 
radiative cooling of electrons should be efficient down to their lowest 
energies. Having from Eqs. (\ref{gammac}) and (\ref{uext}) 
\be \gamma_c \simeq
\frac{r}{4.3 \times 10^{16}\,{\rm cm}} \, 
\frac{1}{(g_u/0.1)(\xi/0.1)(\Gamma/20) L_{d,47}}\,,
\label{nuc}
\ee
one can see that electrons can be cooled down to 
$\gamma_c \sim 1$ when the event is located at $r \lesssim 0.01$ pc.

In such a case, one could expect a bulk-Compton (BC) excess in the X-ray spectra (Begelman \& Sikora 1987). 
At such close proximity to the accretion disk, 
the ERC cooling is dominated by Comptonization of 
the direct accretion disk radiation (Dermer \& Schlickeiser 2002).
Noting that typical energy of external photons at these
distances is of the order $\sim 1$ eV,
one can find the observed energy of the BC feature:
\be
{\rm h}\nu_{\rm BC} \sim \frac{0.4 (\Gamma/20)^2}{(1+z)}\,{\rm keV}\,.
\ee
Such a feature cannot be detected at cosmological distances, unless $\Gamma > 20$.

One might also consider electron cooling at somewhat larger distances,
taking into account the uncertainties of parameters $g_u$ and $\xi$.  
Because high ionization lines, which are produced closer to the black hole than 
the low ionization lines, form much less flattened geometry, the 
value of $g_u$ in the inner BLR can be larger than adopted by us fiducial 
value $0.1$. Also the value of $\xi$ can be larger than 0.1 according to 
some analyses (Kollatschny \& Zetzl 2013).
Noting these uncertainties, one cannot exclude the possibility
that both parameters are underestimated by a factor few, and that 
the distance at which $\gamma_c$ reaches value $\sim 1$ is $\sim 0.1$ pc, 
where external radiation is dominated by the broad emission lines.
However in such a case, energy of the BC feature is predicted to be
located at 
\be
{\rm h}\nu_{\rm BC} \sim \frac{4 (\Gamma/20)^2}{(1+z)} \,~[{\rm keV}]\, .
\ee 
Noting that typical Lorentz factors  implied by ERC(BLR) models are
$\Gamma \sim 15$ (see, e.g., Celotti \& Ghisellini 2008, Table A1)
and that X-ray spectra of most FSRQs show no steepening nor flattening
in the 0.1-2.4 keV band (Sambruna 1997; Lawson \& McHardy 1998), that
prediction seems to contradict with observations.

\subsection{Production of X-ray spectra with $\alpha < 1$ by the SSC process}
\label{sec_SSC}

When the electron injection function at $\gamma > \gi$ has a slope $2<p<3$,
the production of radiation around the spectral component maxima 
(hereafter: spectral peaks) will be dominated by electrons with
energies around either $\gi$ or $\gamma_c$.
Hence, noting that the distance at which both values are equal is
\begin{eqnarray}
r_{\rm ci} &\equiv& r(\gamma_c=\gi) =
\\\nonumber
&=& 2.2 \times 10^{19} \, \left(\frac{\Gamma}{20}\right)
\left(\frac{g_u}{0.1}\right)\left(\frac{\xi}{0.1}\right)
\left(\frac{\gi}{500}\right) L_{d,47} \,[{\rm cm}]\,,
\label{rcb}
\end{eqnarray}
and that cooling break energy increases with the distance from the black hole,
the spectral peaks will be determined by electrons with 
$\gamma \sim \gi$ at distances $r<r_{\rm ci}$, and by electrons with
$\gamma \sim \gamma_c$ at distances $r>r_{\rm ci}$.
In case of $p>3$ the spectral peaks will be associated with the injection energy
over all distances.

\subsubsection{Association of the SSC peak with the average electron energy
injection}
\label{sec_SSCinj}

Electrons injected with the sharp, low energy break at $\gi$ and the 
slope $p >2$ at larger energies  produce SSC spectral component with 
the peak at
\be
\nu_{\rm ssc,i} \simeq \gi^2 \nu_{\rm syn,i} =  c_B B' \gi^4 \Gamma\,,
\label{sscp1}
\ee
where $c_B \simeq 3.7 \times 10^6\,{\rm Gauss}^{-1} s^{-1}$.

Assuming that magnetic field intensity decreases with the distance like
$B' \propto 1/r$,
and is scaled according to the relation $u_B' = u_{\rm ext}'/q$, where
$q \equiv L_{\gamma}/L_{\rm ir/opt}$,
we obtain 
\be
B' = \frac{\Gamma}{r}\,  \sqrt{\frac{2 g_u \xi L_d}{ \pi c q}}\,.
\label{BB}
\ee
The requirement that X-ray spectra are produced by the SSC process with
$\alpha_x < 1$ up to tens of keV's (Ajello et al. 2009)
implies the location of the SSC peak in the range $10-100$ keV.
Then, Eqs. (\ref{sscp1}) and (\ref{BB}) give
\begin{eqnarray}
\gi &\simeq& 660 \, r_{\rm pc}^{1/4} \,
\left(\frac{{\rm h} \nu_{\rm ssc,i}}{30{\rm keV}}\right) \,
\left[\frac{(q/10)}{(g_u/0.1)(\xi/0.1)L_{d,47}}\right]^{1/4}\times
\nonumber\\
&&\times\left(\frac{\Gamma}{20}\right)^{-1/2} \, .
\end{eqnarray}
where $r_{\rm pc} \equiv r/1$ pc.
For such injection energies, for $n_e=n_p$ and $\sigma < 1$,
Eq.(\ref{inj}) implies the energy dissipation
efficiency $\eta_{\rm diss} \sim 42\%$ at $r \sim 1$ pc.

\subsubsection {Association of the SSC peak with the cooling break}
\label{sec_SSCcool}

For $r > r_{\rm ci}$, the SSC peak is located at
\be
\nu_{\rm ssc,peak} = \nu_{\rm ssc,c} = c_B B \Gamma \gamma_c^4\,.
\label{sscc}
\ee
This equation, together with Eqs. (\ref{rcb}), (\ref{nuc}) and (\ref{BB}),
gives the distance at which the SSC peak associated
with the cooling break will be located at ${\rm h}\nu_{\rm ssc,c} = 30$ keV:
\begin{eqnarray}
r_{\rm ssc,c} &\simeq& 1.3 \times 10^{20}\,
\left(\frac{\Gamma}{20}\right)^{2/3}
\left[\left(\frac{g_u}{0.1}\right)\left(\frac{\xi}{0.1}\right)L_{d,47}\right]^{7/6}
\times\nonumber\\&&\times
\left(\frac{q}{10}\right)^{1/6}\,
\left(\frac{{\rm h}\nu_{\rm ssc,c}}{30 {\rm keV}}\right)^{1/3}\,[{\rm cm}]\,;
\end{eqnarray}
and the value of the cooling break energy:
\begin{eqnarray}
\gamma_c &\simeq& 3.0 \times 10^3 \,\left(\frac{\Gamma}{20}\right)^{-1/3} 
\left[\left(\frac{g_u}{0.1}\right)\left(\frac{\xi}{0.1}\right)L_{d,47}\right]^{1/6}
\times\nonumber\\&&\times
\left(\frac{q}{10}\right)^{1/6}
\left(\frac{{\rm h}\nu_{\rm ssc,c}}{30 {\rm keV}}\right)^{1/3}\,.
\end{eqnarray}


\subsection{Can the observed X-ray spectra be produced co-spatially
with $\gamma$-rays and optical radiation?} 
\label{sec_onezone}

Correlations of optical and $\gamma$-ray variabilities,
often observed in FSRQs, suggest a co-spatiality of their emission zones.
In the framework of the ERC model for $\gamma$ rays,
this allows to use two observables --- $q=L_{\gamma}/L_{\rm IR/opt}$
and $w=\nu_{\rm\gamma,peak}/\nu_{\rm IR/opt}$ ---
to estimate the location of that zone.
This comes from relations
\be
q = \frac{L_{\gamma}}{L_{\rm IR/opt}} \simeq \frac{u_{\rm ext}'}{u_B'}\,
\label{qq}
\ee
and
\be
w = \frac{\nu_{\rm\gamma,peak}}{\nu_{\rm IR/opt}}
=\frac{\nu_{\rm erc}}{\nu_{\rm syn}}\,.
\label{ww}
\ee
These equations imply two possible regions where a co-spatial ERC and 
synchrotron emission may take place:
one within the BLR, where ERC is seeded by broad emission lines; 
and one at distances larger by factor $\nu_{\rm BLR}/\nu_{\rm HD} \sim 30$, 
where ERC seeding is provided by IR photons from the dusty torus 
(Sikora et al. 2009 and refs. therein). 
Such a `degeneracy', in the sense of the same synchrotron and ERC 
spectral peak locations at two different distances, 
is broken for the SSC component.  
This is because the production of spectral peaks at these two distances 
involves different -- by a factor of $\sqrt{30}$ -- energies of electrons.  
And noting that $\nu_{\rm ssc} \propto {\gamma}^4 B'$ where $B' \propto 1/r$, 
we expect very different locations of the SSC peaks. While in the outer 
dust region domain the SSC peak can be produced at $\sim 30$ keV 
energies, within the BLR the SSC peaks are located at 
$\sim \left(\sqrt{30}\right)^4/30 = 30$ times lower energies, i.e. at $\sim 1$ keV.

Hence, within the BLR the X-ray spectra with $\alpha_x < 1$
cannot be produced by the SSC process operating co-spatially
with the gamma-ray emission via ERC.
In order to reproduce the entire broad-band spectra with IR/optical
and $\gamma$-rays produced in BLR, the X-rays must originate
either from the SSC process located at distances
$r > r_{\rm ci} > r_{\rm BLR}$ (see \ref{sec_SSCcool}),
or from the low energy tail of the ERC component
located at distances $r < r_{\rm BLR}$ (see \ref{sec_tail}).
%
%
Since the time scales of the X-ray variations are usually longer than
those of the $\gamma$-ray variations,
larger distances of the X-ray production are more likely.
Furthermore, $\gamma$-rays produced co-spatially with X-rays
at $r > r_{\rm ic}$  can explain the VHE radiation observed
in some luminous blazars
(3C 279: Aleksi\'c et al. 2011a;
PKS 1222+216: Aleksi\'c et al. 2011b; Tanaka et al. 2011;
PKS 1510-089: H.E.S.S. Collaboration 2013; Barnacka et al. 2013),
which in the BLR is expected to be strongly absorbed
(see Tavecchio \& Ghisellini 2012 and refs. therein).
At the same time, a contribution from the synchrotron emission
produced at $r > r_{\rm ic}$ to sub-mm radiation can explain variations
in this band on the times scales of weeks (Sikora et al. 2008),
and can suppress the variability amplitude in the FIR band (Nalewajko et al. 2012).
Such a two-zone model can be verified by searching for correlations between
sub-mm, X-ray, and TeV variabilities.

\section{Discussion}
\label{sec_disc}

The origin of X-ray emission in FSRQs was long ago recognized
as an important probe of the jet physics with implications for
the nature of the gamma-ray emission. In the EC scenario,
the X-ray emission would probe the low-energy
(transrelativistic) electrons, and there was a hope that
X-ray spectra would reveal a so-called bulk-Compton component
produced by a population of cold electrons
(Begelman \& Sikora 1987). Lack of strong observational
evidence of the bulk-Compton feature (although see
Kataoka et al. 2008, Ackermann et al. 2012) places 
constraints on the e$^+$e$^-$ pair content, according to which $n_e/n_p < 30$  
(Sikora \& Madejski 2000).

As we demonstrate in Section \ref{sec_pairs}, the pair content is further
constrained, down to $n_e/n_p <$ a few, if noting that the
electrons and protons are strongly coupled, according to
PIC simulations of relativistic shocks (Sironi \& Spitkovsky 2011).
The main part of this paper is devoted to the spectral consequences
of such a coupling.
The $e-p$ coupling implies extremely hard low-energy tails of the 
electron injection function, and therefore
extremely efficient radiative cooling is required to explain 
the X-ray spectra with slopes clustered around $\alpha_x \sim 0.6$. As 
shown in Section \ref{sec_tail}, such spectra can be reproduced only
at very small distances from the black hole, where seeding of the ERC process is 
dominated by direct radiation from the accretion disk.

Another option is that the production of X-rays is dominated 
by the SSC process (Kubo et al. 1995).  
As we demonstrate in Section \ref{sec_SSCinj}, 
consistency of the theoretical SSC spectral slopes with the observed 
X-ray indices is achievable provided that electrons contributing 
to the spectral peaks have energies $\gamma_{\rm peak} > 500$. 
Such energies are too large to explain 
the location and separation of the synchrotron and ERC peaks in the BLR, 
but are of the same order as those predicted by the models which locate  
the blazar zone on distance scales of the dusty torus.  
Hence, on these larger scales 
the X-ray spectra can be produced co-spatially with the optical emission and 
$\gamma$ rays, explaining the occassionally observed 
correlation between all these spectral components, 
as observed e.g. in 3C 454.3 (Bonnoli et al. 2010; Vercellone et al. 2011; 
Wehrle et al. 2012). 

In Section \ref{sec_SSCcool}, we consider the case of X-ray production 
at distances where the cooling break energy, $\gamma_c$, becomes larger than 
the average injection energy, $\gi$.  
For $\gi > 500$, this corresponds to $r > 10$ pc.  
If the X-rays produced by the SSC in this region 
dominate over the X-rays produced by the SSC process in the BLR, 
while the synchrotron and ERC components are produced in BLR, 
this can explain the lack or very limited correlation 
of the X-ray variations with the optical and $\gamma$-ray variations,
as observed e.g. in 3C 279 (Hayashida et al. 2012). 

If the SSC component really dominates the X-ray emission,
and at the same time the EC component dominates
the gamma-ray emission, these components must intersect
at some intermediate photon energy. If this transition
takes place in the hard X-ray band, it can be easily
probed by NuSTAR. Some indications of such a transition can be seen in the
spectrum  of 3C454.3, where the \emph{Swift}/BAT points are located somewhat
above extrapolation of XRT data (see Fig. 4 in Bonnoli et al. 2011). 
This seems to be also consistent
with the Suzaku data (Abdo et~al. 2010b) and 
 INTEGRAL data (Vercellone et~al. 2011).
Even if NuSTAR will not detect any spectral break,
it will still place strong constraints on the low-energy
end of the ERC component.

\section{Conclusions}
\label{sec_concl}

Strong electron-proton coupling in relativistic jets assures that
a large fraction of the dissipated energy is tapped by electrons.
This, and very low radiative efficiency of hadrons injected
with spectral indices $p > 1$, strongly favor the leptonic radiation models
of the luminous blazar spectra.
The SSC origin of X-rays with the observed X-ray slopes 
$\alpha_X < 1$ implies a very large average electron injection energy.
Together with the condition of the high efficiency of energy dissipation,
that implies a rather modest electron-positron pair content.
A co-spatial production of the dominant SSC component
with the observed synchrotron and ERC components
is possible on parsec distances,
where ERC is produced by Comptonization of hot dust radiation.
Lack of correlation of the X-ray variability with the optical and
$\gamma$-ray variations may suggest the origin of X-ray emission at
$r \gtrsim 10$ pc,
with the synchrotron and $\gamma$ rays produced in the BLR.

\acknowledgments 

M.S. thanks Dr. T. Hovatta for providing a copy of the review talk presented 
at EWASS13, which helped to draft the paper.  
We acknowledge financial support by
the Polish NCN grant DEC-2100/01/B/ST9/04845, the NSF grant AST-0907872, the NASA ATP grant NNX09AG02G.

K.N. was supported by NASA through Einstein Postdoctoral Fellowship
grant number PF3-140112 awarded by the Chandra X-ray Center,
which is operated by the Smithsonian Astrophysical Observatory
for NASA under contract NAS8-03060.


\appendix

\section{Energy density of radiation from planar external sources}

\subsection{Radiation energy density}

Intensity from an element of the axisymmetric {\it optically thin} planar 
source is
\be
I_{\rm ext}= j_{\rm ext}s = j_{\rm ext}\frac{2h}{\cos{\theta_{\rm ext}}} =
\frac{1}{8\pi^2 R\cos{\theta_{\rm ext}}}\frac{\partial L_{\rm ext}}{\partial R}\,,
\ee
where $2h$ is the thickness of the source, and $R$ is the radius of the planar
source ring.

Intensity from an element of the axisymmetric {\it optically thick} planar 
source, neglecting limb darkening, is
\be
I_{\rm ext} = \frac{F_{\rm ext}}{\pi} = \frac{1}{4\pi^2 R}
\frac{\partial L_{\rm ext}}{\partial R}\,.
\ee

These intensities differ by factor $2\cos\theta_{\rm ext}$,
and they can be written together as
\be
I_{\rm ext} = 
\frac{f_d(\theta_{\rm ext})}{8\pi^2 R\cos\theta_{\rm ext}}
\frac{\partial L_{\rm ext}}{\partial R}\,,
\label{I-planar}
\ee
where for optically thin source $f_d(\theta_{\rm ext})=1$,
and for optically thick source $f_d(\theta_{\rm ext})=2\cos\theta_{\rm ext}$.

Energy density of radiation from the planar source  in 
the jet co-moving frame at a distance $r$ is then equal to
\be
u_{\rm ext}'(r) = \frac{1}{c}\int{I_{\rm ext}' d\Omega_{\rm ext}'}
= \frac{1}{c}\int{\frac{I_{\rm ext}}{{\cal D}_{\rm ext}^2} d\Omega_{\rm ext}}
= \frac{\Gamma^2}{4\pi c}\int_{R_1}^{R_2}
\frac{(1-\beta\cos\theta_{\rm ext})^2 f_d(\theta_{\rm ext})}{r^2 + R^2} 
\,\frac{\partial L_{\rm ext}}{\partial R}\,dR\,.
\label{uextexact}
\ee
In the above, we used the following relations:
\begin{eqnarray}
{\cal D}_{\rm ext} &=& \frac{1}{\Gamma(1-\beta\cos{\theta_{\rm ext}})}\,,
\\
d\theta_{\rm ext} &=& \frac {\cos\theta_{\rm ext}\,dR}{\sqrt{r^2+R^2}}
= \frac{r\,{\rm d}R}{r^2 + R^2}\,,
\\
{\rm d}\Omega_{\rm ext} &=& \sin\theta_{\rm ext}\,{\rm d}\theta_{\rm ext}\,
{\rm d}\phi_{\rm ext}
= \frac{rR\,dR\,d\phi_{\rm ext}}{(r^2 + R^2)^{3/2}}\,.
\end{eqnarray}

\subsection{Planar external sources}

\subsubsection{Broad-line-region and dusty torus}

It is increasingly accepted that neither broad-line regions (BLR) nor dusty tori (DT) have spherical geometry.
More likely, they are both stratified and flattened,
and as such they can be much better approximated by planar,
vertically thin rings enclosed within distance ranges
$[R_{\rm BLR,1};R_{\rm BLR,2}]$ and $[R_{\rm DT,1};R_{\rm DT,2}]$,
respectively.
Luminosity produced within a ring of thickness ${\rm d}R$
located at distance $R$ from the black hole,
is 
\be 
\frac{\partial L_{\rm ext}}{\partial R}\,{\rm d}R
= \xi_{\rm CF} L_d C_R R^{-s}\,{\rm d}R\,,
\label{Lext}
\ee
where $\xi_{\rm CF}$ is the covering factor of the central source
contributed by the ring
(in general, it can depend on $R$ but we assume it is constant), and
\be
C_R = \left\{
\begin{array}{ccc}
\frac{s-1}{1/R_{1}^{s-1} - 1/R_{2}^{s-1}} & {\rm for} & s \neq 1 \\
\frac{1}{\ln(R_{2}/R_{1})} & {\rm for} & s = 1
\end{array}
\right.\quad.
\ee
Here, we assumed that the external source is optically thin
(in the sense that there is no shadowing of clouds by other clouds),
and that the luminosity has a power-law distribution with $R$.

Recent models of DT and BLR (Elitzur \& Shlosman 2006; 
Elitzur 2008; Czerny \& Hryniewicz 2011), 
BLR reverberation and stratification studies (Peterson 1993; 
Gaskell et al. 2007; Bentz et al. 2006; Kaspi et al. 2007; Mor \& Netzer 2011) 
and interferometric MIR measurements of DT
(Kishimoto et al., 2011) suggest that
$R_{\rm BLR,1} \sim 0.1 R_{\rm sub}$,
$R_{\rm BLR,2} \sim R_{\rm DT,1} = R_{\rm sub}$,
and $R_{\rm DT,2} \sim 10 R_{\rm sub}$,
where 
\be
R_{\rm sub} = 1.6 \times 10^{-5} L_d^{1/2} (1800\,{\rm K}/T_{\rm sub})^{2.8}\,
\ee
(Mor \& Netzer 2011) and $T_{\rm sub}$ is the sublimation 
temperature of the graphite grains (its exact value depends on the grain size). 

The BLR spectra, $\nu L_{\rm BLR,\nu}$ have a peak around $10$~eV, and low-energy
tails which can be approximated by a power-law function with an index
$\alpha_{\rm BLR} \sim 0$ (Poutanen \& Stern 2010). Using monoenergetic
approximation, we adopt ${\rm h} \nu_{\rm BLR} = 10$~eV.
The DT spectra are $\nu L_{\rm IR,\nu} \sim {\rm const}$ in the wavelength range  
$2 - 20 {\rm \mu m}$ and decrease fast
beyond that range (see, e.g., Fig. 4 in Nenkova et al. 2008, and
Fig. 1 in H\"onig et al. 2011).
They can be roughly reproduced by assuming that
$\nu_{\rm DT} = 10^{14} (R_{\rm DT,1}/R)^{\alpha_{\rm DT}}$~Hz, 
where $\alpha_{\rm DT} = \log(\nu_2/\nu_1)/\log(R_{\rm DT,1}/R_{\rm DT,2})$.

Our choice of indices $s$ is: $s_{\rm BLR} = 2$,
in order to have the peak of BLR luminosity close 
to $R_{\rm BLR,1}$,
where contribution from strongest lines Ly${\alpha}$ and C~IV is maximal;
and $s_{\rm DT}=1$ in order to provide $\nu L_{\rm DT,\nu} \sim {\rm const}$ 
for the relation $\nu_{\rm DT} \propto R^{-\alpha_{\rm DT}}$ assumed above.

\subsubsection{Accretion disk}

The total rate at which energy is dissipated in a Keplerian accretion disc
in a ring between $R$ and $R+{\rm d}R$ at a distance $R \gg R_{\rm BH}$ is
\be
\frac{\partial L_{\rm d}}{\partial R}\,{\rm d}R = 
\frac{3GM_{\rm BH}\dot M}{2 R^2}{\rm d}R\,,
\ee
where $\dot M = L_{\rm d}/(\eta_{\rm d} c^2)$.

\subsection{Geometrical correction $g_u$}

We calculate the geometrical correction term
\be
g_u \equiv \frac {4 \pi r^2 c u_{\rm ext}'}{\xi L_{\rm d}\Gamma^2}\,
\label{eq:gu}
\ee
for planar external radiation sources, and present it in Fig. 1.

\begin{figure}
\includegraphics[scale=0.7,angle=-90]{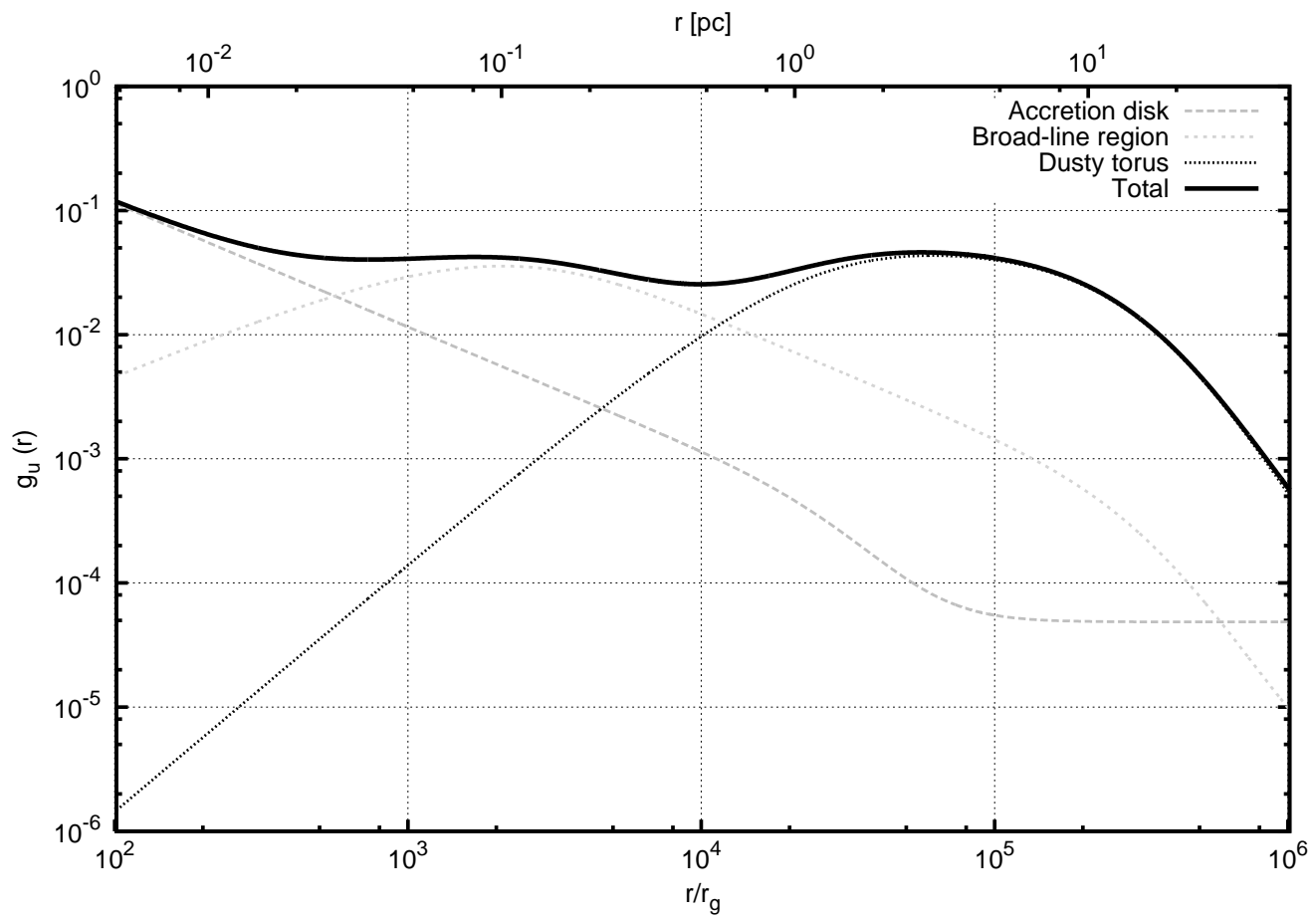}
\caption{Geometrical correction factor $g_u$ as defined in Eq.~\ref{eq:gu} for external radiation planar sources: accretion disk, broad-line region and dusty torus. Parameters used in calculations: $M_{BH}=10^9 M_{\odot}$, $\dot{M}_{d}c^2/L_{Edd} = 10$ and $\eta_{diss}=0.5$.}
\label{fig:1}
\end{figure}

Since the geometries of external radiation sources are not
perfectly planar, the real values of $g_u$ are expected to be 
a bit larger than presented in Fig.~\ref{fig:1}.
We consider $g_u \sim 0.1$ to be reasonable order of its magnitude.


\end{document}